\documentclass[aps,prl,reprint,groupedaddress,showpacs]{revtex4-1}
\usepackage{graphicx}
\usepackage{bm}
\usepackage{hyperref}
\usepackage{color}
\usepackage{amssymb}
\usepackage{color}

\begin{document}


\title{Fiber-optical switch controlled by a single atom}

\author{Danny O'Shea}
\author{Christian Junge}
\author{J\"urgen Volz}
\author{Arno Rauschenbeutel}
\email[]{Arno.Rauschenbeutel@ati.ac.at}
\affiliation{Vienna Center for Quantum Science and Technology, Atominstitut, Vienna University of Technology, 1020 Vienna, Austria}

\date{\today}

\begin{abstract}
We demonstrate highly efficient switching of optical signals between two optical fibers controlled by a single atom. The key element of our experiment is a whispering-gallery mode bottle-microresonator, which is coupled to a single atom and interfaced by two tapered fiber couplers. This system reaches the strong coupling regime of cavity quantum electrodynamics (CQED), leading to a vacuum Rabi splitting in the excitation spectrum. We systematically investigate the switching efficiency of our system, i.e., the probability that the CQED fiber-optical switch redirects the light into the desired output. We obtain a large redirection efficiency reaching a raw fidelity of more than 60\% without post-selection.  Moreover, by measuring the second order correlation functions of the output fields, we show that our switch exhibits a photon number-dependent routing capability.
\end{abstract}

\pacs{42.50.Pq,42.50.Ct,42.60.Da,03.67-.a}

\maketitle

Fiber-optical switches are devices that enable optical signals to be rerouted to different fiber output ports and play a vital role in optical communication networks. Scaling such a device into the quantum domain, where a single quantum system controls the flow of light, would enable the implementation of quantum communication and information protocols with atoms and photons as well as the preparation of non-classical light, useful for interferometric schemes in quantum metrology \cite{Giovannetti2004,Giovannetti2011}.

The physical realization of such a quantum switch requires an enhanced light--matter interaction that can, e.g., be reached by coupling an atom to an optical microresonator. However, this requires to reach the single-atom strong coupling regime, where the so-called critical atom number $N_0=2\kappa\gamma/g^2$ \cite{Ber94}, has to be much smaller than one. Here, $\kappa$ and $\gamma$ are the decay constants of the cavity field and the atomic dipole and $g$ is the single photon--single atom coupling strength. This regime has been investigated  in the optical domain in numerous groundbreaking experiments using high finesse Fabry-P\'erot microresonators \cite{Birnbaum:2005ab,Boozer:2007aa,Wilk:2007aa,Terraciano:2009aa,Kampschulte:2010aa,Volz2011,Ritter:2012aa}. While these experiments clearly demonstrate the high potential of CQED systems for future applications, light absorption and scattering in the mirrors limit the efficiency of coupling light into and out of the resonator to typically a few tens of percent.

\begin{figure}[h]
 \includegraphics[width=0.45\textwidth]{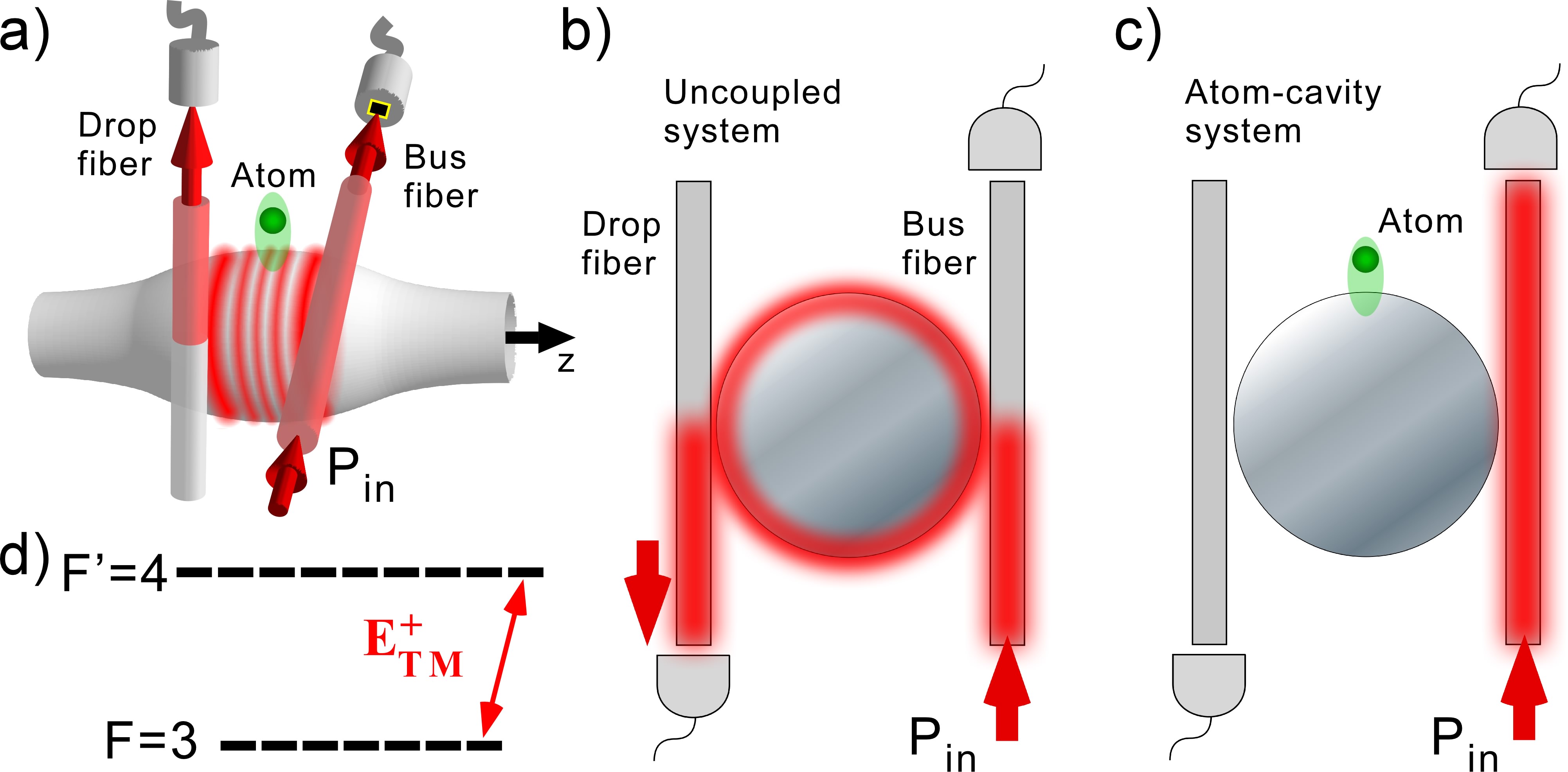}%
 \caption{Simplified experimental setup for the four-port optical switch. (a) Single $^{85}$Rb atoms couple to the TM-polarized evanescent field of a WGM bottle-microresonator. Spectroscopy light couples into the resonator with the bus-fiber and is transferred to the drop-fiber.  Two single photon counters in a Hanbury Brown-Twiss configuration in each output detect the transmitted light. (b,c) The presence or absence of an atom  coupled to the evanescent field controls the flow of light between bus- and drop-fiber. (d) Excerpt from the $^{85}$Rb level scheme with the  Zeeman sublevels relevant for the experiment. \label{Fig:1}}
 \end{figure}

In this context, whispering-gallery-mode (WGM) microresonators combine very high atom-light coupling strengths \cite{Buck2003,Spi04,Lou05} and low coupling losses in the same system. WGM resonators are monolithic dielectric structures, such as microspheres \cite{Bra89} and microtori \cite{Arm03}, in which the light is guided near the surface by continuous total internal reflection \cite{Matsko2006}. The light can be coupled in and out by frustrated total internal reflection with near 100$\%$ efficiency using tapered fiber couplers \cite{Spillane2003}, thereby largely outperforming all other types of optical resonators.

Strong coupling of single atoms and solid state quantum emitters to WGM microresonators has recently been demonstrated in a number of experiments \cite{Aoki:2006aa, Park:2006aa, Srinivasan:2007aa, Dayan:2008aa, Aoki:2009aa, Alton:2010aa, Jun13}. Moreover, using toroidal WGM microresonators, the basic functionality of a turnstile for photons has been implemented \cite{Dayan:2008aa, Aoki:2009aa}: Interfacing the resonator by a single coupling fiber, a weak incident classical light field was split into two counterpropagating streams of bunched and anti-bunched photons, respectively. 

Here, we interface a WGM microresonator by two independent coupling fibers and operate it in an add-drop configuration \cite{Rok04,Poe10}. Passive add-drop filters are used in optical communication for, e.g.,  wavelength-division multiplexing, where a given wavelength channel is unconditionally rerouted. In contrast, the add-drop filter realized in our experiment can be reconfigured by a single $^{85}$Rb atom that controls the switching of light between the two fibers. 

For our studies, we use a novel type of silica WGM microresonator, a so-called bottle-microresonator \cite{Sumetsky2004,Lou05,Poe09}. It is conceptually similar to other WGM microresonators but has the additional advantage of being fully tunable. At the same time, it offers a highly advantageous mode-geometry that enables the simultaneous low-loss coupling to two tapered optical fibers with a nanofiber waist for in/out-coupling of light (see Fig.~\ref{Fig:1}). This renders these resonators true four-port devices, ideally suited for the implementation of highly efficient, narrow-band add-drop filters \cite{Rok04,Poe10}.

The operating principle of our switch is depicted in Fig.~\ref{Fig:1}. The bottle-microresonator in our experiment combines ultra-high quality factors ($Q\approx5 \times 10^7$) with small mode volumes, thereby enabling operation in the strong coupling regime of CQED \cite{OSh11,Jun13}. Two nanofibers, called bus- and drop-fiber, are simultaneously coupled to the resonator mode with coupling constants $\kappa_{\mathrm{bus}}$ and  $\kappa_{\mathrm{drop}}$, respectively. We set the resonator--fiber distances such that the coupling constants for the bus-fiber fulfills the critical coupling condition $\kappa_{\mathrm{bus}}=\sqrt{(\kappa_{i}+\kappa_{\mathrm{drop}})^2+h^2}$ \cite{Rok04}, where $\kappa_i=2\pi\times4.8$~MHz is the intrinsic loss-rate and $h=2\pi\times 1.7$ MHz is the back-scattering rate of the resonator mode. Due to the low backscattering, we can neglect $h$, and the total  decay rate of the resonator field is  $\kappa=\kappa_i+\kappa_{\mathrm{bus}}+\kappa_{\mathrm{drop}}$. 

With these settings, we obtain the situation illustrated in Fig.~\ref{Fig:1} (b), where ideally all light incident through the bus-fiber is resonantly coupled into the resonator and the remaining bus-fiber transmission,  $T^0_{\mathrm{bus}}$, is zero. At the same time, a large fraction $T^0_{\mathrm{drop}}$ of the light  is out-coupled from the resonator and transmitted into the drop-fiber. This situation corresponds to the ON-state of the switch. In order to prepare the OFF-state,  a modification of the resonance frequency of the resonator is required. 
This can be realized in the strong coupling regime by a single atom in the resonator mode, whose presence prevents the build up of the resonator field. In this case, all incident  light remains in the bus-fiber [see Fig.~\ref{Fig:1} (c)]. 

In order to demonstrate this scheme in our experiment, we tune the bottle-microresonator and the probe light into resonance with the atomic transition 5$S_{\mathrm{1/2}}, F = 3 \rightarrow 5P_{\mathrm{3/2}}, F' = 4$ of $^{85}$Rb (wavelength $\lambda = 780$~nm) and set the power of the incident light to $P_{\mathrm{in}}\simeq15-20$~photons/$\mu$s.  We choose the polarization of the probe light in the bus-fiber such that it couples into the TM-polarized resonator mode. In this case, due to the presence of a strong longitudinal electric field component, the clockwise (counterclockwise) propagating resonator mode is nearly perfectly $\sigma^-$- ($\sigma^+$-) polarized, with respect to the quantization axis defined by the resonator fiber [see z-axis in Fig.~\ref{Fig:1}(a)]. In our experiment,  we probe the counterclockwise propagating mode and optically pump the atom into the $F = 3$, $m_F=3$ Zeeman sublevel. As a consequence, the $\sigma^+$-polarized cavity field drives the atomic cycling transition, which effectively leads to the ideal case of a two-level atom that only interacts with a single traveling-wave mode despite the simultaneous existence of two degenerate resonator modes \cite{Jun13}.

\begin{figure}[b]
 \includegraphics[width=0.4\textwidth]{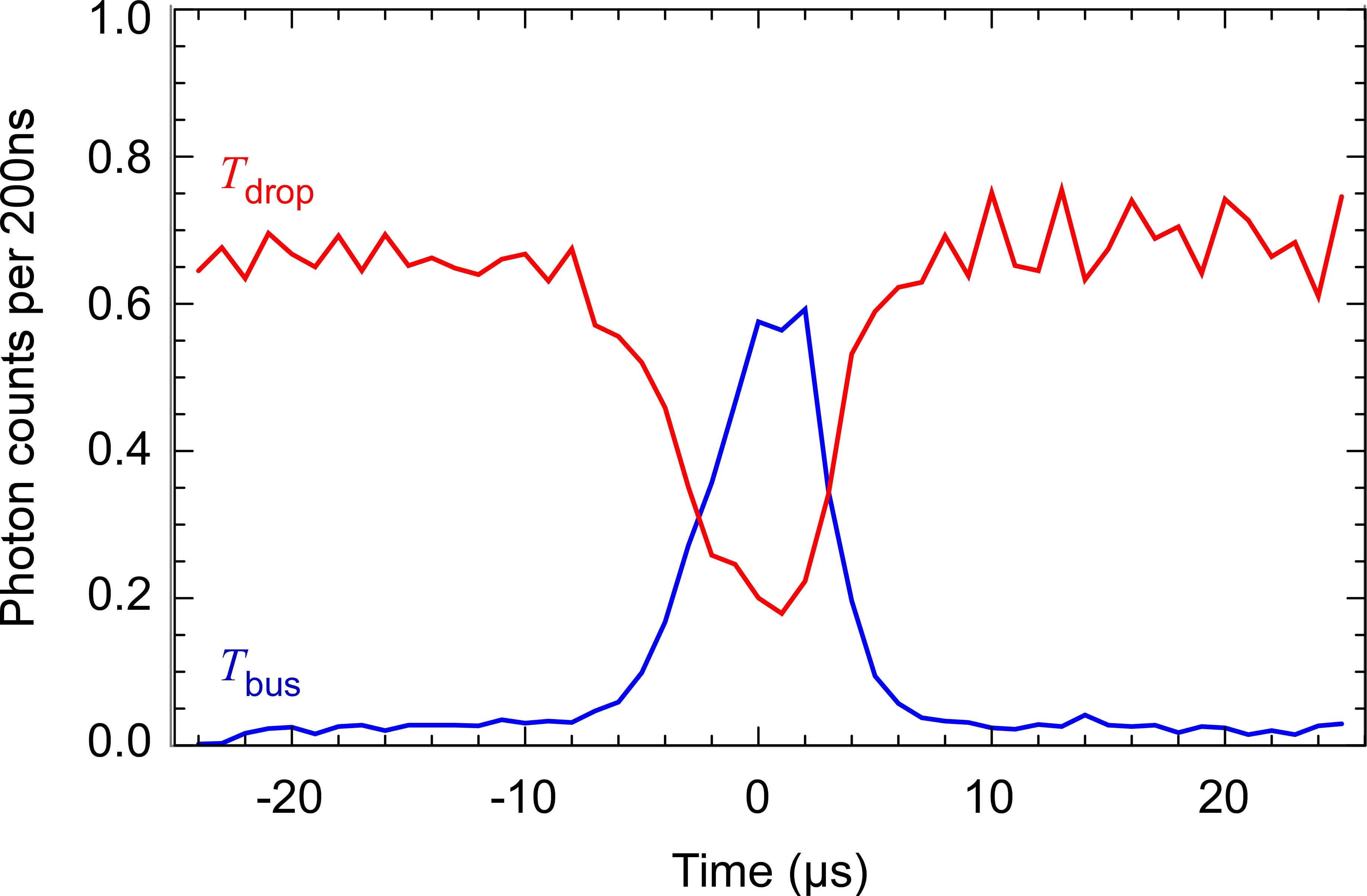}%
 \caption{On-resonance bus-fiber and bus-to-drop-fiber transmission (timebin: 200 ns) during the transit of an atom through the resonator mode, for the same experimental setting as in Fig.~\ref{Fig:3}. The plotted data shows the transmission signal  averaged over 294 single atom transits. The centers of mass of the individual transmission signals are aligned to $t=0$ $\mu$s.\label{Fig:2}}
 \end{figure}

In order to couple atoms to the resonator, our setup is mounted in an ultra-high vacuum chamber in which a 1-cm diameter cloud with $5\times10^7$ laser-cooled atoms is delivered to the resonator using an atomic fountain. Figure~\ref{Fig:2} displays the time-dependent bus- and drop-fiber transmission ($T^{\mathrm{}}_{\mathrm{bus}}$ and $T^{\mathrm{}}_{\mathrm{drop}}$) while an atom passes through the resonator mode. When the atom-resonator coupling becomes significant, the incoming light field  is prevented from entering the resonator and remains in the bus-fiber. We thus observe a concomitant increase (decrease) in bus- (bus-to-drop-) fiber transmission. The temporal width of this transit signal of around 5 $\mu$s is consistent with the expected transit time for an atom with a thermal velocity corresponding to the 5 $\mu$K-temperature of the atom cloud. In order to obtain reproducible experimental conditions on this time scale, a FPGA-based real-time experimental control system detects the presence of an atom in the resonator mode by an increase of the photon count rate in the bus-fiber from around 0.1 to 7 photons per 1.2 $\mu$s  and reacts to this event with around 160 ns response time \cite{OSh11,Jun13}

\begin{figure*}[tbh]
 \includegraphics[width=0.9\textwidth]{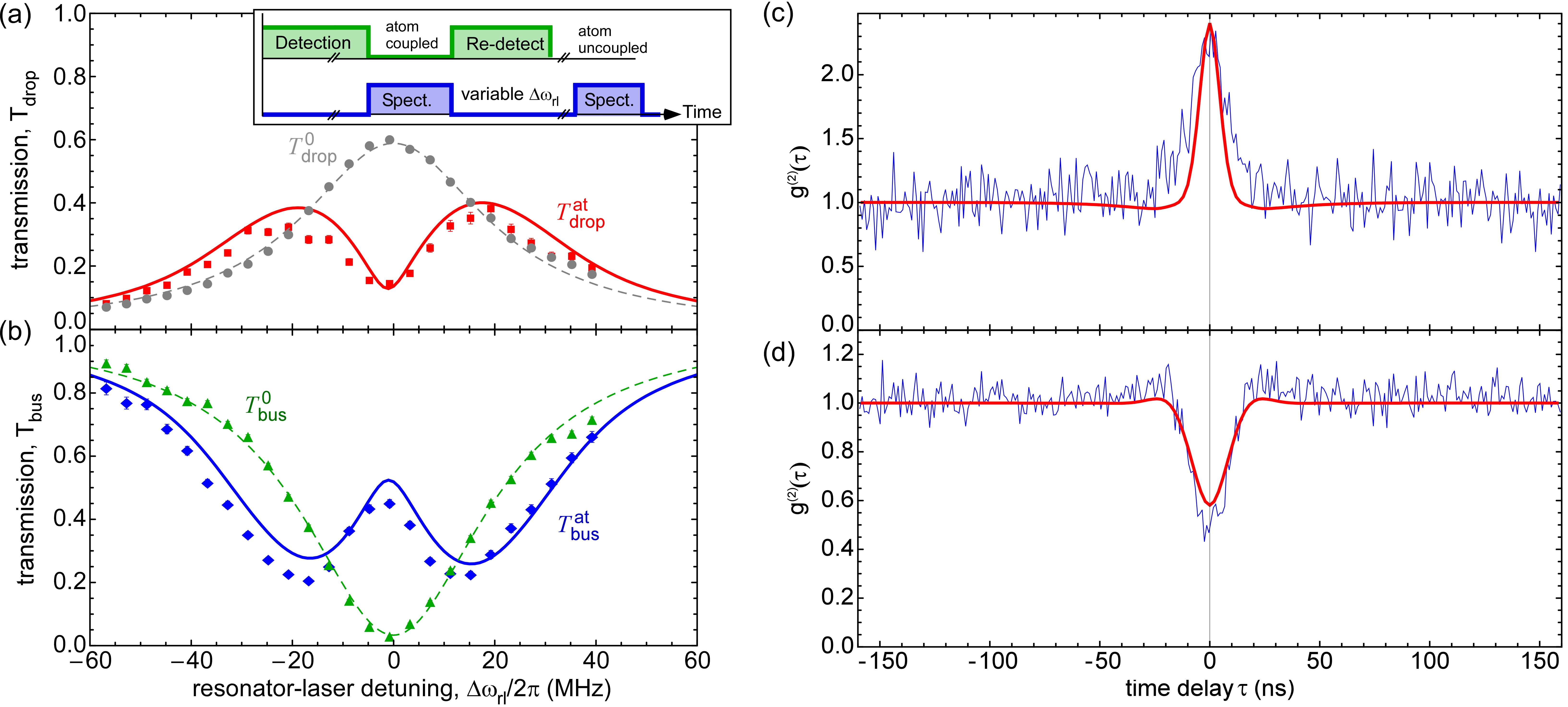}%
 \caption{Normalized transmission spectra of the atom-resonator system  (a) from bus- to drop-fiber $T^{0,\mathrm{at}}_{\mathrm{drop}}$ (grey $\bullet$,  red $\blacksquare$) and (b) through the bus-fiber $T^{0,\mathrm{at}}_{\mathrm{bus}}$ (green $\blacktriangle$, blue $\blacklozenge$). The error bars indicate the $\pm1\sigma$ statistical error and the dashed lines are Lorentzian fits. The inset shows a simplified experimental sequence.  (c) and (d) second order correlation function of the light in the bus- and drop-fiber, respectively, with an atom coupled to the resonator (blue data)  normalized such that $g^{(2)}=1$ for $\tau\gg 1/\kappa$. The solid lines in all panels are obtained from a theoretical fit to the data in (a) and (b). 
  \label{Fig:3}}
 \end{figure*}

We use our real-time control system to perform spectroscopy on the coupled atom--resonator system by scanning the frequency $\omega_l$ of the probe laser across the common resonator and atom resonance frequency $\omega_r=\omega_a$. Figure~\ref{Fig:3} shows the transmission spectra measured at both output ports as a function of the resonator--laser detuning $\Delta\omega_{rl}=\omega_r-\omega_l$. The blue $\blacklozenge$ (red $\blacksquare$) mark the bus- (bus-to-drop-) fiber transmission with an atom coupled to the resonator mode, while the green $\blacktriangle$ (grey $\bullet$) represent the transmission after removing the atom from the resonator mode, see inset Fig.~\ref{Fig:3}. We clearly observe an atom-induced splitting of the transmission spectrum into two distinct peaks, the so-called vacuum-Rabi splitting \cite{Ber94}. This experimentally demonstrates that it is possible to strongly couple single atoms to a whispering-gallery mode resonator in add-drop configuration while maintaining a high on-resonance photon survival probability of $\approx60\%$

In order to infer the atom-resonator coupling strength for this dataset, we numerically solve the Master equation of the Jaynes-Cummings Hamiltonian, taking into account the full Zeeman substructure of the atom as well as the full vectorial description of the evanescent electric field of the two counter-propagating WGM modes (see Ref.~\cite{Jun13} for details). We account for the motion of the atom and the resulting distribution of coupling constants, by solving the Master equation for a set of coupling strengths in the interval $g/2\pi=7.5$~MHz to 30~MHz and subsequently fitting a normally distributed sum of these spectra to the data. For this fit, the mean coupling constant $\bar g$ and the standard deviation $\sigma_g$ are the only free parameters. The spectra obtained in this way are the solid lines in Fig.~\ref{Fig:3} and show good agreement with our experimental data for $\bar{g}/2\pi=15.6$~MHz and $\sigma_g=9$~MHz. The residual difference possibly originates from a misalignment of the magnetic guiding field with respect to the resonator axis which would limit the efficiency of optically pumping.

 The maximum switching contrast  is on resonance ($\Delta\omega_{rl}=0$), where the measurement yields a 10 dB increase in bus-fiber transmission (7 dB decrease in bus-to-drop fiber transmission) from 3$\%$ (58$\%$) in the ON-state to 46$\%$ (12 $\%$) in the OFF-state of the switch.
Figure \ref{Fig:3} also shows the intensity correlation of the light in the  bus- and drop-fiber measured for the same experimental conditions. We observe photon (anti-) bunching in the drop- (bus-) fiber, in good agreement with the theoretical prediction calculated for the coupling strength distribution obtained from the vacuum Rabi splitting. 
This demonstrates that our atom--resonator system allows photon number-dependent rerouting of light: Due to the high non-linearity of the Jaynes-Cummings Hamiltonian, two photon states are preferentially coupled into the resonator and can exit through the drop fiber, while single photons remain in the bus-fiber\cite{Dayan:2008aa,Kubanek2008}.

\begin{figure}[htb]
 \includegraphics[width=0.45\textwidth]{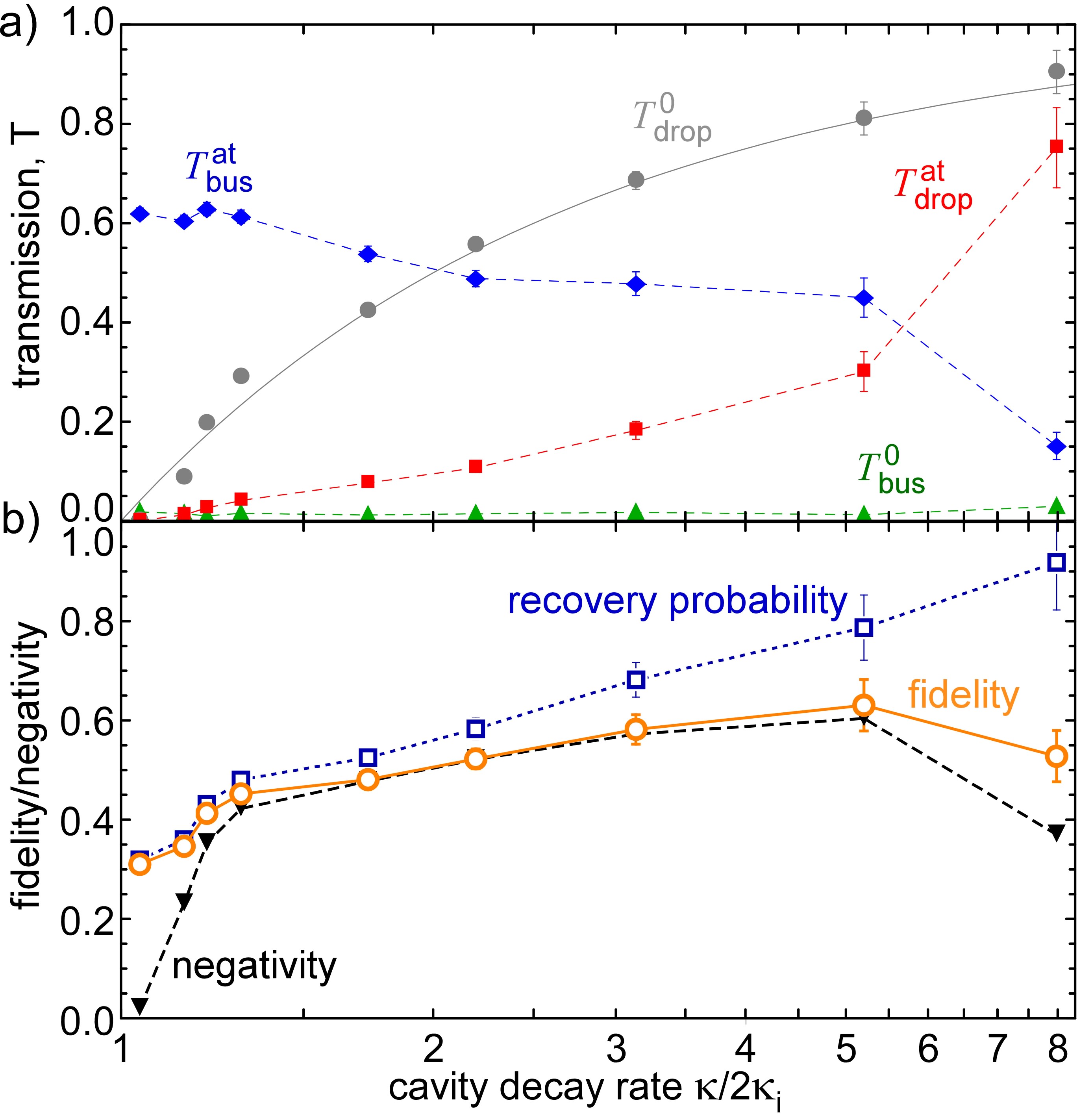}%
 \caption{(a) Measurements of the on-resonance transmission $T^{(\mathrm{at,0})}_{\mathrm{bus}}$ (blue $\blacklozenge$, green $\blacktriangle$) and $T^{\mathrm{(at,0)}}_{\mathrm{drop}}$ (red $\blacksquare$, grey $\bullet$) as a function of the resonator decay rate $\kappa$. The dashed lines are a guide to the eye, and the solid line is given by $T^{\mathrm{0}}_{\mathrm{drop}}=1-2(\kappa_i/\kappa)$. (b) Fidelity of the switch operation (orange $\circ$), probability to recover an incident photon (blue $\square$) and expected negativity for the generation of an entangled Bell state  (black $\blacktriangledown$), inferred from the data in (a). The lines are guides to the eye and the error bars indicate the $\pm1\sigma$ statistical error. \label{Fig:4}}
 \end{figure}
 
In order to find the optimal operating conditions of the switch, we scan the distance between resonator and the drop-fiber while maintaining critical coupling for the bus-fiber. We measure the on-resonance transmission to both output ports as a function of the total cavity decay rate $\kappa$ when an atom couples to the resonator and for the uncoupled case, see Fig.~\ref{Fig:4}~(a). As expected, the effect of the atom decreases with  increasing resonator decay rate $\kappa$ due to the increase of the critical atom number $N_0$ of the atom--resonator system. This can be seen in the monotonous increase (decrease) of the bus-to-drop- (bus-) fiber transmission $T^{\mathrm{at}}_{\mathrm{drop}}$ ($T^{\mathrm{at}}_{\mathrm{bus}}$) with $\kappa$. At the same time, in the uncoupled case, the bus-to-drop-fiber transmission $T^{\mathrm{0}}_{\mathrm{drop}}$ increases monotonously in good agreement with the expected theoretical prediction $T^{\mathrm{0}}_{\mathrm{drop}}=1-2(\kappa_i/\kappa)$ \cite{Spi04} and the remaining transmission through the bus-fiber $T^{\mathrm{0}}_{\mathrm{bus}}$ is approximately zero.

We quantify the switching process by calculating the classical fidelity $\mathcal F=1/2(T^{\mathrm{at}}_{\mathrm{bus}}+T^{\mathrm{0}}_{\mathrm{drop}})$ of the process,  i.e., the raw probability that a single input photon will be routed from the bus-fiber into the correct fiber port, without correcting for photon loss. This fidelity is shown in Fig.~\ref{Fig:4}~(b) and is maximal  for  resonator linewidths between $\kappa\simeq30$--50~MHz. In this range, we find a remarkably constant bus-fiber transmission of $T^{\mathrm{at}}_{\mathrm{bus}}\simeq45-55$\% despite a large increase in the cavity decay rate. For optimum settings of the resonator--fiber coupling, we obtain a fidelity of $\mathcal F=0.62$ and, at the same time, a probability of 79\%  to recover an incident photon after the switching operation. The main limit of this fidelity originates from averaging over different coupling strengths of the atom. In comparison, other error sources as, e.g., photon background counts can be neglected.
We note that the maximum fidelity is not achieved in the strong coupling regime but at the crossover to the regime where the atom--light coupling rate $g$ is smaller than the resonator decay rate $\kappa$, but still dominates over the spontaneous emission rate $\gamma$. Thus, the underlying switching mechanism is robust against experimental variations in $\kappa$ as long as the condition $g^2/\kappa\gamma>1$ is satisfied.  
Due to intrinsic losses $\kappa_i$, the switching operation is asymmetric, and using the switch in reverse direction with the drop-fiber as input port would yield a different fidelity. However, for optimum settings, the coupling rates $\kappa_{\mathrm{bus}}=2\pi\times25$ MHz, $\kappa_{\mathrm{drop}}=2\pi\times20$ MHz are very close, thus yielding approximately the same fidelity for both directions.

A multi-port optical switch that is controlled by a single atom is a powerful tool for future quantum applications such as in quantum networks. In particular, the atom could be prepared in an equal superposition between two internal ground states, e.g., the two hyperfine ground states $F=2$ and $F=3$ of $^{85}$Rb, where only one of them interacts with the resonator. In this case, after the interaction of the switch with an incident photon, the atom--photon system will ideally end up in a maximally entangled Bell-state. Thus, a quantum switch facilitates the deterministic entanglement between two initially independent quantum systems. 
Assuming a coherent interaction between the incident light and the atom-resonator system --- which should be possible using weak input light powers and slowly varying pulse envelopes ($\sim0.1$ $\mu$s for our experimental parameters) --- we calculate the expected performance of our system for the deterministic generation of matter-light entanglement. Under this assumption,  one obtains the same fidelity as  measured for the classical case. To better quantify the amount of entanglement that  can be produced,  we calculate the corresponding negativity \cite{Vid02} of the final atom--photon state for each measurement, see Fig.~\ref{Fig:4}~(b). Even including the photon loss, we obtain a maximum value of around 0.6.

In summary, we demonstrate highly efficient switching of optical signals, where a single atom coupled to a bottle-microresonator controls the output port of an incident light field. In the ON-state of the switch, most of the light is redirected to the drop-fiber, while in the OFF-state and for input powers corresponding to much less than one photon within the cavity lifetime, most of the light exits through the bus-fiber. For larger input powers, the switch in the OFF-state exhibits a photon number-dependent routing capability, where the incoming stream of photons is sorted into single photons and pairs in the two output ports.  
The switching fidelity of more than $\mathcal F=0.62$ in conjunction with the low optical losses -- around $79\%$ of the incident photons are recovered -- illustrates the potential of this system for future applications under realistic conditions. 
In particular, the high degree of light--matter entanglement expected for such a system would render it a powerful tool for many fiber-based quantum information and communication applications and would enable the on-demand generation of Schr\"odinger cat-like entangled states containing many photons. 

The expected switching fidelity of our system can be further improved by reducing the motion of the atoms. Trapping the atoms close to the resonator surface using a nanofiber-based dipole trap \cite{Vet10} would strongly reduce the fluctuations in the coupling rate $g$ and would yield a significantly more stable operation of the switch.  Furthermore, experiments show that it is possible to increase the quality factor of the bottle microresonator by at least a  factor of five \cite{Poe09}. With these improvements and assuming realistic trapping conditions, a deterministic generation of matter--light entanglement with a fidelity/negativity of more than 95$\%$ is within reach, thereby enabling a realm of possible applications in quantum science and technology.

\begin{acknowledgments}
We gratefully acknowledge financial support by the European Science Foundation, the Volkswagen Foundation, and the Austrian Science Fund (FWF; SFB FoQuS Project No. F 4017). J.V. acknowledges support by the European Commission (Marie Curie IEF Grant 300392). C.J. acknowledges support by the German National Academic Foundation.
\end{acknowledgments}

\bibliography{Switch_Bib}

\end{document}